\documentclass[twocolumn]{openjournal}
\usepackage{xcolor}
\usepackage{textgreek}
\usepackage{tikz}
\usepackage[utf8]{inputenc}
\usepackage[english]{babel}

\usepackage{hyperref}
\hypersetup{
    unicode,
    colorlinks=true,
    linkcolor=linkcolor,
    citecolor=linkcolor,
    filecolor=linkcolor,
    urlcolor=linkcolor,
}
\usepackage{color,colortbl}
\definecolor{linkcolor}{rgb}{0.0,0.3,0.5}
\usepackage{tensind}
\tensordelimiter{?}
\DeclareGraphicsExtensions{.bmp,.png,.jpg,.pdf}
\usepackage{verbatim}
\usepackage[normalem]{ulem}
\usepackage{orcidlink}
\usepackage{soul}
\newcommand{\astroeg}[1]{\par\smallskip\noindent\textit{Astronomy example.}~#1}

\graphicspath{ {./figs/} }

\usepackage{newtxtext,newtxmath}
\usepackage[T1]{fontenc}

\usepackage{graphicx}	% Including figure files
\usepackage{amsmath, amssymb, graphicx, color, units}

\usepackage{acro}

\usepackage[table]{xcolor}
\usepackage{colortbl}

\newcommand{\CfA}{Center for Astrophysics \textbar{} Harvard $\&$ Smithsonian,
60 Garden St., Cambridge, MA 02138, USA}

\newcommand{\UCSD}{Department of Astronomy and Astrophysics, University of California, San Diego, La Jolla, CA 92093, USA}

\newcommand{\SDSC}{San Diego Supercomputer Center, University of California, San Diego, La Jolla, CA 92093, USA}
\newcommand{\UofT}{David A. Dunlap Department of Astronomy \& Astrophysics, University of Toronto, Toronto, ON M5S 3H4, Canada}
\newcommand{\DI}{Dunlap Institute for Astronomy \& Astrophysics, University of Toronto, Toronto, ON M5S 3H4, Canada}
\newcommand{\USRSE}{The US Research Software Engineer Association, USA}
\newcommand{\UIUC}{University of Illinois Urbana-Champaign, Urbana, IL, USA}
\newcommand{\NCSA}{National Center for Supercomputing Applications, University of Illinois Urbana-Champaign, Urbana, IL, USA}
\newcommand{\FAIR}{FAIR Data Innovations Innovations Hub, California Medical Innovations Institute, San Diego, CA, 92121 USA}
\newcommand{\UCBerk}{University of California Berkeley, Berkeley, CA 94720, USA}
\newcommand{\ISU}{Illinois State University, Normal, IL 61790, USA}
\newcommand{\Tufts}{Tufts University, Medford, MA 02155, USA}  

\begin{document}

\title{A Practical Primer on Software Citation in Research}

\author{Floor S. Broekgaarden$^{1}$\orcidlink{0000-0002-4421-4962}}
% \email{fbroekgaarden@ucsd.edu}
\author{Phil R. Van-Lane$^{1,2,3}$\orcidlink{0009-0009-4567-9946}}
\author{Sasha Levina$^{1}$\orcidlink{0000-0003-1241-7615}}
\author{Ahmed Bello$^{4}$\orcidlink{0009-0005-1692-4413}}
\author{Sandra Gesing$^{5,6}$\orcidlink{0000-0002-6051-0673}}
\author{Daniel S. Katz$^{7}$\orcidlink{0000-0001-5934-7525}}
\author{Bhavesh Patel$^{8}$\orcidlink{0000-0002-0307-262X}}
\author{Pengyin Shan$^{9}$\orcidlink{0009-0009-6309-380X}}
\author{Samantha Teplitzky$^{10}$\orcidlink{0000-0001-7071-332X}}
\author{Simon Thill$^{11}$\orcidlink{0009-0003-6910-6399}}
\author{Peter K. G. Williams$^{12}$\orcidlink{0000-0003-3734-3587}}
\author{Andrea Zonca$^{5,1}$\orcidlink{0000-0001-6841-1058}}

\affiliation{$^{1}$\UCSD}
\affiliation{$^{2}$\UofT}
\affiliation{$^{3}$\DI}
\affiliation{$^{4}$\ISU}
\affiliation{$^{5}$\SDSC}
\affiliation{$^{6}$\USRSE}
\affiliation{$^{7}$\UIUC}
\affiliation{$^{8}$\FAIR}
\affiliation{$^{9}$\NCSA}
\affiliation{$^{10}$\UCBerk}
\affiliation{$^{11}$\Tufts}
\affiliation{$^{12}$\CfA}

\begin{abstract}
Software underlies virtually all modern research, yet citation practices for software remain inconsistent and often inadequate. 
This short primer offers practical guidance in two directions: how to properly cite the software you use in your research, and how to make your own software easy for others to cite. 
We describe a seven-step workflow for citing software and a five-step workflow for making your software citeable, along with the community tools and resources that support each step. 
We also address common questions about what software, and how much, to cite. 
The guidance is field-independent; we use astronomy as a running example throughout, and readers in other disciplines can substitute the equivalent tools and conventions of their own community.
Our goal is to lower the practical barrier to good software citation practice and to support a research culture in which software receives the scholarly credit it deserves.
\end{abstract}

\begin{keywords}
{software citation, open-source software, reproducibility, scholarly communication}
\end{keywords}

\maketitle

%─────────────────────────────────────────────────────────────────────────────
\section{Introduction}
\label{sec:intro}
%% ─────────────────────────────────────────────────────────────────────────────

Software is central to every stage of modern astronomical research: it drives data-reduction and analysis pipelines, powers numerical simulations, and underpins statistical analyses. Yet despite this centrality, software citation in the literature is inconsistent. 
Authors frequently omit citations for key tools,
acknowledge packages only in prose rather than as formal references, or cite
papers \textit{about} software rather than the software itself \citep[e.g.][]{Howison2016seeing,  Bouquin...2020ApJS..249....8B, Du2021softcite, Schindler2021,  Schindler2023, Druskat:2024, teplitzky2025seismica}, depriving developers of the visible, traceable credit they are due for contributions that may span years of collaborative effort \citep[][]{Bouquin...2020ApJS..249....8B}. 

The barriers are practical as much as cultural. 
Researchers may be unsure how to locate the preferred citation for a package, how to format a software entry in a bibliography, or why a persistent identifier matters \citep[][]{Bouquin...2020ApJS..249....8B}. 
Developers, in turn, may not know that their code is citeable or what approaches to take to make their code easy for others to cite.

Before turning to practice, it is worth being clear about what software citation is for. A citation is more than a mention: it is a reference that can be traced back to a work's authors, so that any reuse becomes visible, attributable credit \citep{Niemeyer2016}. 
For software, citation usually serves primarily as this credit mechanism, rewarding developers and helping them demonstrate the reach and value of their work. 
It can also aid reproducibility, but only in part: stating the version of the software you used helps, yet citation is no substitute for code-level practices such as specifying dependencies (a requirements.txt or environment file), containerization, versioning, and clear documentation \citep[][]{Katz2018, Bouquin...2020ApJS..249....8B,  Katz...FAIR...2021arXiv210110883K}.

Prior guidance on software citation, notably the FORCE11 principles \citep{Smith2016}, the Software Citation Checklists for authors and developers \citep{hong:2019-developer-guide, hong:2019-author-guide}, the guide of \citet{Katz2020}, and the FAIR principles for research software \citep{Barker2022}, established the core practices we follow here. 
This primer complements them in four ways: it pairs a seven-step workflow for citing software with a five-step workflow for making software citeable; it maps a set of community tools onto both; it grounds each step in a running astronomy example, updated for current resources and AI-assisted workflows; and it gives concrete, worked guidance on the question these earlier resources largely leave open, namely what and how far down the dependency tree to cite software.

Software-citation practice varies across disciplines, but the underlying principles are general. 
We therefore keep this guidance field-independent and illustrate each step with a running example drawn from astronomy: the field most of the authors work in. 
Where we write \textit{Astronomy example}, readers in other disciplines should be able to map the illustration onto the equivalent tools, registries, and journal conventions of their own community, and we encourage them to use this primer as a customizable baseline.

\begin{figure*}
  \centering
  \includegraphics[width=0.9\textwidth]{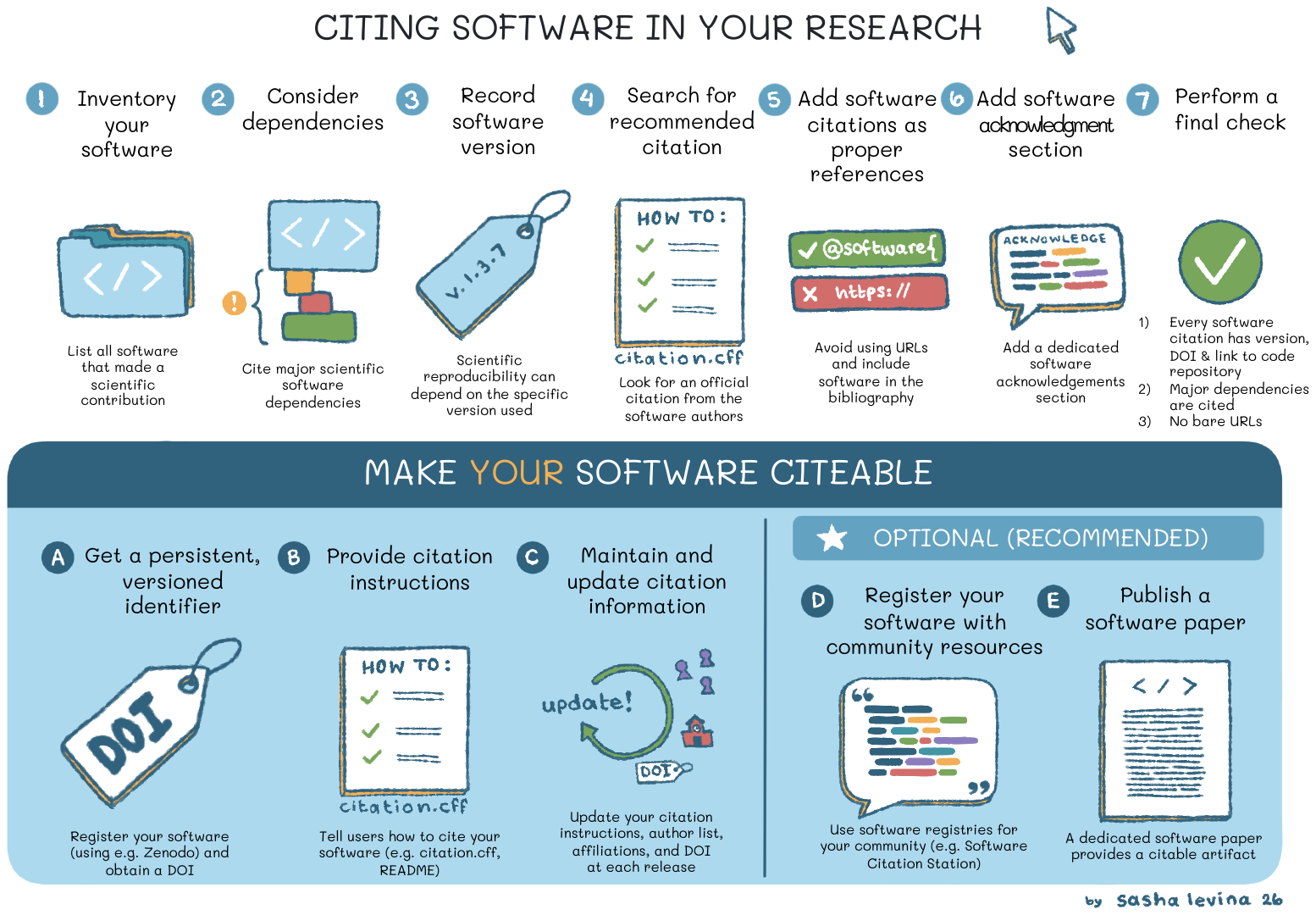}
  \caption{Summary of the recommended steps for citing software (top; Section~\ref{sec:citing}) and for making your own software citeable (bottom; Section~\ref{sec:citeable}). In the bottom panel, the three required steps are shown on the left, and two optional but recommended steps are shown on the right. Figure by Sasha Levina, available for re-use at \citep[][]{Levina:2026-software-citation-graphic}.}
  \label{fig:tools}
\end{figure*}

  We summarize our guidelines in Figure~\ref{fig:tools}.
% This primer addresses both sides of software citation: citing software and making your software citeable.  

%% ─────────────────────────────────────────────────────────────────────────────
\section{Citing Software in Your Research}
\label{sec:citing}
%% ─────────────────────────────────────────────────────────────────────────────

The seven steps below constitute a recommended workflow for properly citing software in a manuscript. 

\begin{enumerate}

%% ── Step 1 ──────────────────────────────────────────────────────────────────
\item \textbf{Inventory your software.}
Go through your analysis scripts, notebooks, and pipelines and compile a list of all software that contributed to the scientific results. 
It is useful to distinguish between \textit{core scientific software} (e.g.~simulation frameworks, inference tools), \textit{supporting analysis packages}, and \textit{infrastructure or workflow tools}. 
For each entry, ask: Did it make a meaningful contribution to
the results? Would citing it help a reader reproduce the analysis?
If the answer to any of these is yes, include it. See
Section~\ref{ap:whatocite} for a fuller discussion of inclusion criteria.
\astroeg{Separate a simulation framework or an inference engine (core
scientific software) from plotting and table-handling utilities (supporting
packages). Both can warrant citation, but the former almost always does.} Tools like LLMs can be used to help with this step. 

%% ── Step 2 ──────────────────────────────────────────────────────────────────
\item \textbf{Consider dependencies.}
Many packages rely on upstream libraries that represent distinct intellectual contributions. 
If a dependency played a significant role in your analysis, even if it is not used directly in the analysis, it may deserve its own citation. 
Tools such as
\href{https://citeas.org/}{CiteAs}  and \href{https://github.com/duecredit/duecredit}{DueCredit} \citep[][]{Halchenko2024duecredit} can help identify recommended citations for packages and their dependencies.
\astroeg{General-purpose packages such as \texttt{numpy} \citep[][]{numpy}, \texttt{scipy} \citep[][]{2020SciPy-NMeth}, \texttt{matplotlib} \citep[][]{Hunter:2007}, and \texttt{astropy} \citep[][]{astropy:2013,astropy:2018,astropy:2022} frequently underlie higher-level tools even when you never import them directly; check whether any of them played a dominant role in your analysis. Tools like \href{https://www.tomwagg.com/software-citation-station/}{Software Citation Station} \citep[][]{Wagg2024} automatically include dependencies.}

%% ── Step 3 ──────────────────────────────────────────────────────────────────
\item \textbf{Record software versions.}
Scientific reproducibility often depends on the exact version of software package used in the analysis: the same code can produce different results across releases as a software package's version changes.  
Citing the correct software version also matters for credit: a package's contributor list may change from one release to the next, so the version you cite determines whose work is recognized.
Record the version of every package on your list. 
You can also record things like the Git commit hash or release tag. 
If a version cannot be recovered, report an approximate date of use together with the repository URL. 
A version-ambiguous citation is still preferable to no citation at all.
\astroeg{In Python, \texttt{pip list} or \texttt{conda list} report installed
versions; most packages also expose a \texttt{\_\_version\_\_} attribute or
respond to \texttt{-{}-version}.}

%% ── Step 4 ──────────────────────────────────────────────────────────────────
\item \textbf{Search for a recommended citation.}
Many packages specify how they prefer to be cited: in their documentation, README, or a machine-readable \texttt{CITATION.cff} \citep[][]{Druskat2021cff} file. 
Follow these instructions; do not substitute an unrelated paper or a bare URL. 
If a package's requested citation itself falls short of best practice (for example, it lacks a DOI or is only a URL), you can supplement or adapt it while preserving the maintainers' intent (see Section~\ref{sec:faq}). 
For machine-learning workflows, also check whether foundational models, datasets, or APIs specify citation or attribution requirements.
\astroeg{Good places to find a package's preferred citation include the \href{https://www.tomwagg.com/software-citation-station/}{Software Citation Station} \citep[][]{Wagg2024}, the \href{https://ascl.net/}{Astrophysics Source Code Library} \citep[ASCL;][]{Allen2016ascl}, \href{https://citeas.org/}{CiteAs}, and the \href{https://www.chorusaccess.org/resources/software-citation-policies-index/}{CHORUS} software-citation-policies index. For tools that cannot be made public, a \href{https://www.rrids.org/}{Research Resource Identifier} \citep[RRID;][]{RRID} supplies a citeable handle.}

%% ── Step 5 ──────────────────────────────────────────────────────────────────
\item \textbf{Add software citations as proper references.}
Formal software citations should appear in the reference list, rather than only in the acknowledgments. 
Create a reference entry in whatever bibliographic format your tooling uses, and include the version number and DOI wherever available.
Avoid citing bare URLs: they are not persistent and do not constitute a formal reference. A \texttt{CITATION.cff} file can be converted to common citation formats automatically: \href{https://pypi.org/project/cffconvert/}{cffconvert} \citep[][]{cffconvert_spaaks_2021_5521767} and \href{https://citation.js.org/}{citation-js} \citep[][]{citation_js} export to BibTeX and other styles, and \href{https://zenodo.org/}{Zenodo} \citep[][]{zenodo} record pages offer one-click exports in several formats. 
 If you write in \LaTeX{}, use the \texttt{@software} entry type (or \texttt{@misc} if your bibliography style does not support it).
 \astroeg{The \href{https://www.tomwagg.com/software-citation-station/}{Software Citation Station} provides ready-made BibTeX entries for commonly used astronomy software. LLMs  can also help identify software used in a workflow and draft candidate BibTeX entries, although these should always be checked against the software’s official citation instructions, documentation, or archived release before use.}
% \astroeg{\href{https://github.com/adrn/makecite}{makecite} \citep[][]{makecite} scans your scripts
% and emits BibTeX for common astronomy packages, a convenient starting point for
% the reference list. The
% \href{https://www.tomwagg.com/software-citation-station/}{Software Citation
% Station} also provides ready-made BibTeX entries.}

%% ── Step 6 ──────────────────────────────────────────────────────────────────
\item \textbf{Add a software acknowledgment section.}
Many journals encourage or require a dedicated software section (separate from the general acknowledgments) listing cited packages with their citation keys.
Consult your target journal's author guidelines. 
Consider also including a software availability statement and a link to your analysis repository or data-release page.
\astroeg{Most astronomy journals expect a short software acknowledgment. The \href{https://journals.aas.org/policy-statement-on-software/}{AAS journals} provide a dedicated \texttt{\textbackslash software\{\}} tag for flagging the code you used, with citations given in the reference list, and \href{https://academic.oup.com/mnras/pages/General_Instructions}{MNRAS} asks that software available online be cited in the reference list and noted in a data-availability statement. 
Consult your target journal's guidelines, and link to your analysis repository or data-release page.}
%% ── Step 7 ──────────────────────────────────────────────────────────────────
\item \textbf{Perform a final check.}
Before submission, verify that (1) every software package that you want to cite appears in the reference list with version information, the link of its code repository and resolvable DOIs; (2) all major scientific dependencies are acknowledged consistently; (3) no bare URLs substitute for formal references. 
Ask yourself: did I give all of the due credit to the software that I used in my analysis?
If not, refine before submitting.
\astroeg{Confirm that \href{https://ascl.net/}{ASCL} \citep[][]{Allen2016ascl} / \href{http://adswww.harvard.edu/}{ADS} \citep{ADS} / \href{https://scixplorer.org/}{SciX}  links and \href{https://zenodo.org/}{Zenodo} \citep[][]{zenodo} DOIs resolve, and that packages you only touched through a pipeline are still acknowledged.}

\end{enumerate}

%%%%%%%%%%%%%%%%%%%%%%%%%%%%%%%%%%%%%%%%%%%%%%%%%%%%%%%%%%%%%%%%%%
%% ─────────────────────────────────────────────────────────────────────────────
\section{Making Your Software Citeable}
\label{sec:citeable}
%% ─────────────────────────────────────────────────────────────────────────────

Making software citeable requires a small amount of extra work but can yield significant long term benefits: more reproducible papers, greater developer credit, and wider community reuse \citep[][]{Barker2022}. 
Three steps are essential; two further steps increase discoverability and long-term impact.

\subsection*{Required steps}

\begin{enumerate}

%% ── Step 1 ──────────────────────────────────────────────────────────────────
\item[A.] \textbf{Get a persistent, versioned identifier.}
A digital object identifier (DOI) is the most reliable way to make a specific version of your software formally citeable \citep[][]{Barker2022}. 
We recommend \href{https://about.zenodo.org/}{Zenodo} \citep[][]{zenodo}, the free CERN-supported repository: it accepts arbitrary files, guarantees long-term preservation, and mints DOIs with proper metadata. 
Zenodo supports \href{https://zenodo.org/help/versioning}{DOI versioning}: a concept DOI represents the software as a whole, while version-specific DOIs point to individual releases, allowing users to cite exactly the version they ran.
Give each release a clear version number, ideally following semantic versioning \citep[][]{PrestonWerner2013}, so that version-specific DOIs and citations map cleanly onto identifiable releases.
GitHub users can enable the Zenodo integration directly from their repository settings, so every tagged release is automatically archived and assigned a DOI.
Note, however, that this DOI is issued only \emph{after} the release is archived, so the release's own \texttt{CITATION.cff} cannot yet carry its version-specific DOI, leaving that metadata a step behind, which runs counter to keeping citation information current (see the maintenance step below). 
Recording the concept DOI (above) is a simple partial fix; to keep version-specific DOIs in your metadata, use a release workflow that reserves the DOI first, writes it into \texttt{CITATION.cff} \citep[][]{Druskat2021cff} and \texttt{codemeta.json} \citep[][]{Jones2017codemeta}, and only then publishes the release. Tools such as \href{https://codefair.io}{Codefair} \citep[][]{codefair_portillo_2026_19261487} automate this.
\href{https://figshare.com}{Figshare} and institutional repositories offer comparable functionality. For software that cannot be made publicly available, a Research Resource Identifier (RRID) \citep[][]{RRID} provides a persistent identifier without requiring open access.
\astroeg{The \href{https://help.zenodo.org/docs/github/}{GitHub--Zenodo integration} is the least-effort route for most repositories or just uploading software to Zenodo directly; the ASCL \citep[][]{Allen2016ascl} additionally provides a citeable handle for codes that predate a DOI (see below).}

%% ── Step 2 ──────────────────────────────────────────────────────────────────
\item[B.] \textbf{Provide citation instructions.}
Once your software has a DOI, tell users explicitly how to cite it. 
The preferred approach is a \texttt{CITATION.cff} \citep[][]{Druskat2021cff} file, a structured, machine-readable metadata file placed in the root of your repository. 
GitHub automatically detects it and adds a ``Cite this repository'' button to the sidebar. 
A \texttt{codemeta.json} \citep[][]{Jones2017codemeta} file serves the same purpose in a complementary standard. 
Tools including \href{https://pypi.org/project/cffconvert/}{\texttt{cffconvert}} \citep[][]{cffconvert_spaaks_2021_5521767}, the \href{https://codemeta.github.io/codemeta-generator/}{CodeMeta generator}, and the \href{https://citation-js.github.io/cff-generator/}{citation-js CFF generator} \citep[][]{citation_js} simplify creating and validating these files. 
At minimum, include a citation section in your README pointing to the Zenodo record or paper.
\astroeg{Point your README's citation section at the Zenodo record and, if you have one, the paper representing the code.}

%% ── Step 3 ──────────────────────────────────────────────────────────────────
\item[C.] \textbf{Maintain and update citation information.}
Citation metadata should evolve with the software: update your \texttt{CITATION.cff}, author list, affiliations, and DOI link at each release.
This is especially important for collaborative projects with growing or changing contributor bases \citep{2026arXiv260817159S}. Automated release pipelines can handle this end-to-end and serve as a practical template for best practice.
\astroeg{The \href{https://github.com/tardis-sn/release-pipeline-template}{TARDIS release-pipeline template} \citep[][]{TARDIS_release_pipeline_template} automates metadata, archiving, and DOI minting at each tagged release, and provides a good starting point to adapt.}

\end{enumerate}

\subsection*{Optional but recommended}

\begin{itemize}

\item[D.] \textbf{Register your software with community resources}. Most science fields have custom software registries. Check with your community. 
\astroeg{Register with the \href{https://ascl.net}{ASCL} and the
\href{https://www.tomwagg.com/software-citation-station/}{Software Citation Station}. ASCL entries are indexed by ADS and receive a persistent identifier that can itself serve as a citation target.}

\item[E.] \textbf{Publish a software paper.}
A dedicated software paper provides a citeable artifact familiar to most researchers. 
The \href{https://joss.theoj.org}{Journal of Open Source Software} \citep[JOSS;][]{JOSS...2017arXiv170702264S} specializes in exactly this: it peer-reviews software together with the article, and makes the software citeable through conventional paper-based workflows.
Note that JOSS mints a single DOI for a specific version of the paper, rather than version-specific DOIs, so a Zenodo record remains valuable alongside a JOSS paper. 
Other journals publish software-focused articles, including ``living'' papers updated across software generations, while JOSS may have multiple papers for the same software as it moves through major versions, each with its own DOI.
\astroeg{Beyond JOSS, venues such as \textit{RAS Techniques and Instruments} \citep[RASTI;][]{rasti_tennyson_ras_2022} and the AAS journals publish software- and methods-focused articles.}

\end{itemize}

%%%%%%%%%%%%%%%%%%%%%%%%%%%%%%%%%
%% ─────────────────────────────────────────────────────────────────────────────
\section{Discussion}
\label{sec:discussion}

%% ─────────────────────────────────────────────────────────────────────────────
\subsection{Frequently asked questions}
\label{sec:faq}

\textbf{Should I cite every piece of software I use?}
No. Be thoughtful rather than exhaustive. Prioritize software that made a meaningful intellectual or practical contribution to your results and that has a formal, citeable identifier. 
General-purpose infrastructure (operating systems, compilers, standard text editors) rarely needs to be cited.
See Section~\ref{ap:whatocite} for a fuller discussion of where to draw the line.
 
\textbf{What if I cannot find the DOI or preferred citation for a package?}
First, search the package documentation, README, and repository for any stated citation information; usable citation details sometimes exist even without a DOI. If none is given, construct the citation yourself from four elements \citep{hong:2019-author-guide, Smith2016}: who (the project name, or the individual if authorship is clear), when (the release date of the version you used, or your access date otherwise), what (the software name and version), and where (a persistent identifier, preferring a DOI to a bare URL). In BibTeX this is a \texttt{@misc} entry with those fields plus a version tag or commit hash. A bare URL is acceptable only as a genuine last resort.

\textbf{What if a package's requested citation is nonstandard, incomplete, or points to an unrelated paper?}
Because software-citation conventions are not yet standardized, requested citations vary in quality: some omit a DOI or other persistent identifier, some are bare URLs, and some ask you to cite a loosely related paper. 
Follow the maintainers' stated preference where you can, since it is how they wish to receive credit, but you may adapt or supplement it: add the Zenodo or archival DOI if one exists, note the version or commit you used, and record an access date for a bare URL. 
If the requested reference is genuinely broken (a dead link or the wrong work), flag it to the maintainers via a repository issue.

\textbf{Who should be an author on your software?}
Software authorship need not match the author list of any associated paper: the people credited in a software citation are those who made a meaningful contribution to the software itself, and such contributions extend beyond code to design, documentation, testing, packaging, and sustained maintenance. 
Because the contributor roster changes from release to release, revisit the author list (the \texttt{CITATION.cff} \texttt{authors} field and the Zenodo creators) at each release rather than fixing it once. There is no universal rule, so make the decision collaboratively and transparently, distinguishing the citation authors from the wider set of contributors you acknowledge; when in doubt, err toward inclusion. 
Many projects record everyone in a \texttt{CONTRIBUTORS} file and
elevate a subset to citation authors.

\textbf{Is a software availability statement the same as citing the software?}
No. A software (or code) availability statement records the software's name, its repository location, and any conditions on access, and complements the reference list rather than replacing it \citep[][]{hong:2019-author-guide}. 
The formal citation gives credit and appears among your references; the availability statement tells readers where the code lives and how to obtain it. Some publishers require both.

\textbf{Do I need to license my software for it to be cited?}
No. Citation and licensing are separate concerns: you can cite software regardless of its license, and the license has no bearing on whether you should cite it \citep[][]{hong:2019-author-guide}. 
For your own software, however, add an explicit open-source license \citep[][]{hong:2019-developer-guide}: without one, the legal default is that others cannot reuse your code, so a citable release can still be an unusable one. Choose a license that fits your community's needs (\href{https://choosealicense.com/}{choosealicense.com} is a useful starting point).
 
%% ─────────────────────────────────────────────────────────────────────────────
\subsection{Takeaways and the Road Ahead}
\label{sec:takeaway}
%% ─────────────────────────────────────────────────────────────────────────────
 
Software citation is not an afterthought: it is a core component of reproducible science and a mechanism for sustaining the community infrastructure that research depends on. 
The workflows described here require only modest effort but yield concrete benefits: more reproducible papers, more visible software, and more sustainable development teams.
 
Looking ahead, the growing use of LLMs and AI-assisted tools raises new questions for citation and attribution practice. 
When an AI tool contributes materially to an analysis or generates code used in a publication, the norms for disclosure are still evolving. 
We encourage researchers to apply the same transparency they would to any other computational tool: be explicit about what was used, how, and to what extent. 
Emerging frameworks such as voluntary Model Influence Statements \citep[][]{Shan_Model_Influence_Statement_2026} aim to provide structured vocabulary for this, and community standards are expected to mature rapidly. 
The core principle, however, is unchanged: \textit{cite with care, and give credit where it is due.}

\subsection{On What and How Much to Cite}
\label{ap:whatocite}
There is no single, universally agreed-upon answer to which software should be cited, nor to how extensively one should trace the dependency tree. 
This ambiguity mirrors the broader challenge in scholarly writing of deciding which prior works merit citation: in both cases, citation practice requires judgment rather than a fixed rule.
 
A useful guiding principle is to \textit{cite with care}. 
Consider whether a piece of software made a meaningful intellectual or practical contribution to
your work; whether it is formally citeable (i.e., it has a DOI or a stated preferred citation); and whether citing it helps a reader reproduce your analysis or gives appropriate credit to its developers. 
Software citations can directly benefit the individuals and teams behind the tools: improving visibility, funding prospects, and career recognition. 
This impact is especially significant for community-developed scientific tools maintained by small teams or early-career researchers, for whom citation counts may be primary evidence of software impact.
 
The same considerations apply to dependencies. 
Underlying libraries represent real intellectual contributions, but citing every layer of a software stack is often impractical and may not align with journal norms or reviewer expectations.
A reasonable approach is to cite the highest-level packages you interacted with directly, plus any foundational dependencies that played a distinctive or dominant role in your results.

As a concrete astronomy example, consider a gravitational-wave population-synthesis study that runs \texttt{COMPAS} for the binary evolution, processes the output with \texttt{numpy} and \texttt{scipy}, computes cosmological quantities with \texttt{astropy}, and makes figures with \texttt{matplotlib}. Here \texttt{COMPAS} is core scientific software and is always cited. \texttt{astropy} is cited as well, since its cosmology and units routines shape the results directly rather than merely formatting them. \texttt{numpy} and \texttt{scipy} are the judgment calls: try to cite them when a specific routine drives a result, such as a \texttt{scipy} interpolation or optimization scheme central to the analysis, and otherwise treat them as general infrastructure and decide for yourself whether to cite them. \texttt{matplotlib} is a supporting package that many authors cite, at little cost. The software dependencies beneath these packages are usually not cited individually. Versions are recorded for every software package on the list and the COMPAS code itself, so the environment stays reproducible even where a citation is not warranted.

Ultimately, software citation is an evolving practice, with norms varying across fields, collaborations, and journals. Rather than prescribing a fixed threshold, we advocate for a thoughtful and intentional approach: \textit{cite with care, and give credit where it is due.} Be transparent about the computational tools underlying your results, prioritize reproducibility, and make an effort to cite the software used in your work.
When genuinely uncertain, err toward inclusion: the cost of an unnecessary citation is far lower than the cost of unrecognized labor.

\section*{Acknowledgments}
The authors acknowledge support for the 2026 Software Citation Workshop from the NASA TWSC program under award number 316634-00001. 
PV, SL, and FSB acknowledge support from NASA HPOSS grant 80NSSC25K7555 under award number 316592-0000. FSB and SL acknowledge support from NSF 22-624 Astronomy and Astrophysics Research Grants under award number 2606407. PV acknowledges support from a Postgraduate Scholarship - Doctoral (PGS-D) Award [PGS D - 589175 - 2024] from the Natural Sciences and Engineering Research Council of Canada (NSERC).
AZ acknowledges support from NSF Mid-Scale Research Infrastructure-2 grant number 2153201 (PI: Mark Devlin, University of Pennsylvania) for the Advanced Simons Observatory; Simons Foundation Award 457687.
This research has made use of the Science Explorer, funded by NASA under Cooperative Agreement 80NSSC21M00561 as well as the Astrophysics Data System, funded by NASA under Cooperative Agreement 80NSSC21M0056. Software citation information aggregated using \texttt{\href{https://www.tomwagg.com/software-citation-station/}{The Software Citation Station}} \citep{software_citation_station, Wagg2024}.
We developed, drafted and revised all content and text of this paper ourselves, but used Claude Opus 5 to edit for text and grammar typos throughout (similar to advanced spell check), after which we iterated/edited the text ourselves further. We also used it to brainstorm on Figure~\ref{fig:tools} design, but final figure was hand drawn and fully created by Sasha Levina.

\bibliographystyle{aasjournal}

% You should give the same name for your .bbl as your main .tex
% since it is a requirement for posting on ArXiv.
\bibliography{software-primer}

\end{document}